\documentclass[aps,prl,twocolumn,showpacs,preprintnumbers,amsmath,amssymb,floatfix]{revtex4-2}

\usepackage{graphicx}        
\usepackage{dcolumn}         
\usepackage{bm}              
\usepackage{amsmath}         
\usepackage{amssymb}
\usepackage{xcolor}          
\usepackage{hyperref}        
\usepackage{comment}    
\usepackage{float}      
\begin{document}

\title{The fractal dimension of Brownian dynamics in liquids}

\author{Michael C. Thumann}
\affiliation{Department of Physics, The University of Texas at Austin, Austin, TX 78712, USA}
\author{Jason Boynewicz}
\affiliation{Department of Physics, The University of Texas at Austin, Austin, TX 78712, USA}
\author{Giuseppe Procopio}
\affiliation{Facolt\`{a} di Ingegneria Civile and Industriale, DICMA,  Universit\`{a} di Roma La Sapienza, via Eudossiana 18, 00184, Roma, Italy}
\author{Massimiliano Giona}
\affiliation{Facolt\`{a} di Ingegneria Civile and Industriale, DICMA, Universit\`{a} di Roma La Sapienza, via Eudossiana 18, 00184, Roma, Italy}
\author{Mark G. Raizen}
\affiliation{Department of Physics, The University of Texas at Austin, Austin, TX 78712, USA}
\email[corresponding author:]{thumann@utexas.edu}
\date{\today}

\begin{abstract}

The classical Einstein-Langevin theory of Brownian motion assumes a memoryless thermal bath, establishing a universal fractal dimension of $d_v=3/2$ for the velocity fluctuations of a particle. In this Letter, we demonstrate experimentally and theoretically that fluid-inertial memory effects fundamentally redefine the fractal scaling of these fluctuations. In analyzing highly resolved measurements of Brownian microspheres in liquids, we show that the non-Markovian hydrodynamic thermal noise establishes a distinct velocity fractal dimension of $d_v = 7/4$. Coupled with theoretical analysis of non-equilibrium short-time dynamics and the initial scaling of the velocity autocorrelation function, this result establishes the non-equilibrium universality class of Brownian motion in fluid
media possessing a finite non-vanishing density.

\end{abstract}
\maketitle
Brownian motion stands as one of the cornerstones of modern physics, defining our understanding of natural phenomena controlled by fluctuations across a multitude of studies and disciplines \cite{bm1,bm2,bm3,bm4,bm5}. Although Brownian motion theory is long-standing and well-established, the coupled contributions of 
highly resolved experiments \cite{hres1,hres2,hres3,hres4,btr},
and of more refined hydrodynamic
approaches \cite{hydro1,hydro2,hydro3,hydro4,chowhermans,gionapof}
allow for the revision of canonical findings. In this Letter, we show that the fractal dimension of  Brownian velocity 
fluctuations, controlled by hydrodynamic  fluid-inertial
effects \cite{landau,kim},  equals $7/4$ and not $3/2$ as in the Einstein-Langevin theory (see Appendix G). This universality class (App. E) of Brownian dynamics is supported also by the experimental analysis of short-time non-equilibrium
behavior of Brownian particles in liquids. 

The concept of universality class requires some discussion. The classical interpretation of it refers to the long-term properties of Brownian motion, and it is 
completely defined by the fractal dimension  $d_x$ of the
particle trajectories $x(t)$ vs $t$, $d_x=3/2$.
A different result for the fractal dimension is found at small time scales, 
as the  particle position $x(t)$ is, up to an additive constant (the initial
position),
the primitive function of the velocity. Consequently,
  its graph is necessarily
a Lipschitz-continuous function of time, and thus $d_x=1$.
Nevertheless, the statement $d_x=3/2$  reflects  in a compact way the long-term
statistical properties of Brownian motion, as specified by
the Einstein relation for the mean square displacement
$\langle (x(t) - x(0))^2 \rangle \sim 2D \,  t$, where $D$
is the particle diffusivity, justifying and supporting all
the parabolic models of transport theory \cite{onsager,green1,green2}.
In this framework, different universality classes of
random kinematics can be defined  in terms of  deviations
from the Einstein relation, in the
form of anomalous diffusive motion $\langle (x(t) - x(0))^2 \rangle \sim  t^\alpha$,
whether subdiffusive ($\alpha < 1$), superdiffusive ($1< \alpha  < 2$),
or ballistic ($\alpha=2$) \cite{walk}, and
the long-term universal
feature of Brownian motion compactly entails its equilibrium
properties, and is completely specified by the exponent $d_x=3/2$.

Conversely, the fine structure of the particle
velocity fluctuations controls the short-time non-equilibrium properties 
\cite{btpg}, see also \cite{btr,duplat,wm}.
This represents the other facet of the universal behavior
of Brownian motion and, as discussed below, all its properties
are subsumed by the value of the fractal dimension
of the particle velocity fluctuations. 

In order to describe the experimental conditions of highly resolved Brownian
experiments, we considered a spherical micrometric particle of mass $m$ and radius $R_p$ in a still
liquid at constant temperature $T$, confined
 by the action of an optical trap, acting on the
particle as a harmonic potential (App. D).
We adopt  a nondimensional formulation where  the nondimensional
time $\tau=t \, \eta/m$ is defined with respect to
the momentum relaxation time $t_{\rm rel}=m/\eta$, 
$\eta=6 \, \pi \, \mu \, R_p$ being
 the Stokesian friction factor, and $\mu$ 
the fluid viscosity,  the
velocity $v(t)$ is rescaled via the characteristic
thermal velocity  $v_{\rm th}=\sqrt{k_B \, T/m}$,
where $k_B$ is the Boltzmann constant,
to define the normalized quantity $u(t)=v(t)/v_{\rm th}$,
from which it follows that the position is normalized
with respect  to the characteristic thermal
displacement length $v_{\rm th} \, t_{\rm rel}$.
In this 
setting, the  equilibrium value
of the nondimensional velocity variance is $\langle u^2 \rangle_{\rm eq}=1$.

The Einstein-Langevin theory  of Brownian motion \cite{langevin}
is defined by the stochastic dynamics
\begin{eqnarray}
\frac{d x(\tau)}{d \tau} & = & u(\tau) \nonumber \\
\frac{d u(\tau)}{d \tau} & = & - u(\tau)- \kappa_s \, x(\tau) + 
\sqrt{2}\, \xi(\tau)
\label{eq1}
\end{eqnarray}
where  $\xi(\tau)= d w(\tau)/d \tau$ is the distributional derivative of a 
Wiener process $w(\tau)$, and  $- \kappa_s \, x(\tau)$ 
the harmonic force stemming
from the optical trap ($\kappa_s=k_s \, m/\eta^2$ is the
nondimensional spring constant of the trap
characterized by the physical spring constant $k_s$ [N/m]).
The process defined by eq. (\ref{eq1}) is customarily referred to
as an Ornstein-Uhlenbeck process.

In the Einstein-Langevin approach, the dynamics are Markovian and the particle's velocity is a stochastic process possessing the fractal dimension $d_v=3/2$. This dimension 
stems from the singular structure of the uncorrelated Wiener
process $w(t)$  \cite{falconer},
that, dictated by the fluctuation dissipation theorem, defines the white thermal
force. While the thermal force is a property of the equilibrium thermal bath, these equilibrium fluctuations govern the relaxation of unthermalized initial states \cite{onsager}. Therefore, the non-equilibrium universality
class of the Einstein-Langevin Brownian motion is characterized
by the fractal dimension $d_v=3/2$.

Experimental works and theoretical analyses \cite{hres2,hres3,hres4,
hydro1,hydro2,hydro3,hydro4}
have shown, however, that eq. (\ref{eq1}) fails to interpret some relevant features
of Brownian dynamics in liquid, such as the scaling
of the velocity autocorrelation function.
In Newtonian liquids, such as water or acetone at room temperature, 
an accurate description of the microparticle hydromechanics involves 
the Stokesian friction
and the fluid inertial effects (Basset  force) leading to
the following dynamic equation replacing the second eq. (\ref{eq1})

\begin{multline}
\frac{d u(\tau)}{ d \tau}= - u(\tau)- \kappa_s \, x(\tau) \\ - \int_0^\tau
 k(\tau-\theta)
\, \left ( \frac{d u(\theta)}{ d \theta} +u(0) \, \delta(\theta) \right ) \, d \theta
+ R(\tau)
\label{eq2}
\end{multline}
where the kernel $k(\tau)$ obtains, in the
strictly incompressible case, the expression 
\begin{equation}
k(\tau) = \frac{b}{\sqrt{\tau}} \,,  \qquad b=\sqrt{\frac{9 \, \rho}{2 \, \pi \, \rho_p}}
\label{eq3}
\end{equation}
where $\rho$, and $\rho_p$ are  the fluid
and the  particle densities, respectively.

The kernel $k(\tau)$ in eqs. (\ref{eq2})-(\ref{eq3}) is referred to as the Basset kernel and depends on the fluid density $\rho$,
$k(\tau) \sim \sqrt{\rho}$:  it is therefore related to the fluid inertia. The quantity $R(\tau)$ is
a stochastic process accounting for the thermal/hydrodynamic
fluctuations, the statistical properties of which  derive from
the Kubo theory (and specifically from the
second fluctuation/dissipation relation) \cite{kubo1,kubo2}.

There are two important {\em caveats} associated with eqs. (\ref{eq2})-(\ref{eq3}).
In eq. (\ref{eq2}) the hydrodynamic added mass does not appear,
as it is an effect characterizing  strictly incompressible fluids
occurring as a consequence of the paradox of an infinite
speed of sound associated with the solenoidal nature of the fluid
velocity field. The added mass  vanishes if fluid compressibility
is considered \cite{chowhermans}.  
Therefore, the functional form eq. (\ref{eq2})
explicitly accounts for the solution of this paradox.
Moreover, it follows from eq. (\ref{eq3}) that the Basset kernel
displays a singularity at $\tau=0$. This singularity is
the consequence of another paradox of infinite velocity of propagation
\cite{gionapof},
specifically that of the shear stresses, that disappears once a non-vanishing
shear relaxation time  is accounted for \cite{procopiogiona}. For water
or acetone at room temperature, the stress relaxation time is on the order of
a picosecond \cite{ruocco}, and consequently  
the viscoelastic departure from the 
singular Basset
kernel is significant at time scales that are $10^{-4}$-$10^{-7}$ smaller
than the particle momentum relaxation time.
These observations indicate that for timescales well above the stress
relaxation time, the singular Basset kernel can be accurately used
in the theoretical analysis. Conversely, the  simulation
of Brownian trajectories (including fluid-inertial effects),
and specifically the representation of the thermal-hydrodynamic
force $R(\tau)$ entering eq. (\ref{eq2}) and satisfying the
Kubo fluctuation-dissipation relation should necessarily account for
the boundedness of $k(\tau)$ at $\tau=0$ \cite{gpp1}.

Boynewicz et al. \cite{btr} have recently performed experimental measurements
and  scaling analysis of the early dynamics of the second-order
moment of the position variable $m_{xx}(\tau)= \langle x^2(\tau) \rangle$
in liquids, starting from the initial conditions $u(0)=0$, obtaining
for $m_{xx}(\tau)$ the  scaling
\begin{equation}
m_{xx}(\tau) \sim \tau^{5/2} = \tau^\varphi
\label{eq4}
\end{equation}
The corresponding result for gases (i.e., in the realm of validity
of the Einstein-Langevin theory, see App. G.2) is $m_{xx}(\tau) \sim \tau^3$ as shown
experimentally by Duplat et al. \cite{duplat}.
A thorough analysis of  the initial moment dynamics
associated with the  non-equilibrium
conditions  $u(0)=0$ has been developed in \cite{btpg}, 
where it is shown that
the value of the exponent $\varphi$ entering
eq. (\ref{eq4}),  equal to $5/2$ in liquids and $\varphi=3$ in
gases (in the range of timescales considered in the experiments in \cite{duplat}),  	
is associated with  the regularity properties of the velocity fluctuations.

The local fractal dimension  $d_v$, or equivalently 
the H\"older exponent $H_v$ of  the velocity fluctuations,
is considered here as the characteristic quantity assessing the class of universality of
Brownian dynamics in different fluid media in non-equilibrium conditions. Specifically, the exponent $\varphi$ is related to the H\"older exponent
of velocity fluctuations by the expression \cite{btpg}

\begin{equation}
\varphi = 2 + 2 \, H_v
\label{eq5}
\end{equation}
that in turn implies, following Mandelbrot's geometric formalism \cite{mandelbrot1, tricot}, that the fractal dimension of the realizations of Brownian particle velocity equals
\begin{equation}
d_v= 2 - H_v = \frac{6-\varphi}{2}
\label{eq6}
\end{equation}
For $\varphi=5/2$, one obtains $d_v=7/4$.

\begin{figure}
\includegraphics[width=8cm]{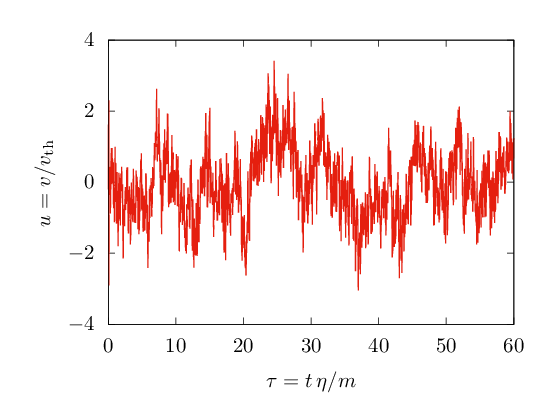}
\caption{Normalized velocity  $u(\tau)$ vs $\tau$ of a barium titanate glass microsphere in acetone.}
\label{Fig1}
\end{figure}

A direct analysis 
of $d_v$ is performed in this Letter by
considering a barium titanate glass microsphere ($R_p=3.4$ $\mu$ m, $\rho_p=4200$ 
kg/m$^3$) in acetone ($\rho=790$ kg/m$^3$, $\mu=0.32$ mPas)
at constant temperature $T=293$ K. For  details
on the experimental system see App. D and \cite{btr}.
Figure \ref{Fig1} depicts  a portion of the  nondimensional velocity
signal $u(\tau)$ of a microsphere in acetone.

Classically, an estimate of $d_v$ follows from length-resolution
analysis \cite{higuchi}, where one
considers the scaling of the normalized length $L_u(\Delta \tau)$ of
the velocity realization $u(\tau)$ vs $\tau$ for $\tau  \in [0,\tau_{\rm max}]$, $\tau_{\rm max} \simeq 67$, as a function
of the yardstick size $\Delta \tau$,

\begin{equation}
L_u(\Delta \tau)= \frac{1}{\tau_{\rm max}} \sum_{i=1}^{N(\Delta \tau)} 
\sqrt{\Delta \tau^2
+  \Delta u_i^2}  \, ,
\label{eq7}
\end{equation}

with $\tau_{\rm max} = N_T(\Delta \tau) \, \Delta \tau$, $\Delta u_i= u(\tau_i+\Delta \tau)-u(\tau_i)$, $\tau_i= i \, \Delta \tau$,
$i=1,\dots, N_T(\Delta \tau)+1$, and analogously for $L_v(\Delta t)$ with
respect to $v(t)$.
When considering the 
scaling of $L_u(\Delta \tau)$, in the small $\Delta \tau$ limit, or
$L_v(\Delta t)$ for small $\Delta t$,
one recovers the velocity fractal dimension $d_v$,
\begin{equation}
L_u(\Delta \tau) \sim  \Delta \tau^{1-d_v}.
\label{eq7a}
\end{equation}

While $L_v(\Delta t)$ serves to evaluate the fractal dimension, its form mixes physical 
units, coupling the time resolution with the stochastic fluctuations which introduces 
an analytical limitation for physical signals. It is the  square root in the 
Pythagorean form that makes the subtraction of independent noises (e.g. laser noise) impractical. This applies also to $L_u(\Delta \tau)$. To resolve these issues, we introduce the \emph{scale-dependent absolute slope} defined as 
\begin{equation}
    S_v(\Delta t)=E\big[|\Delta v|\big]/\Delta t.
\end{equation}
or equivalently  $S_u(\Delta \tau)$ for the nondimensional velocity $u(\tau)$.
When applied to a Brownian particle's velocity time-traces, $S_v(\Delta t)$ has the units of acceleration. Because the increment $\Delta v$ is a zero-mean Gaussian variable with variance $\sigma^2(\Delta t)$, $S_v(\Delta t)$ can be evaluated via the probability density function for $\Delta v$
\begin{equation}
S_v(\Delta t) = \frac{1}{\sqrt{2\pi}\sigma}\int_{-\infty}^{\infty} \left| \frac{y}{\tau} \right| \exp\left[\frac{-y^2}{2 \sigma^2}\right]dy = \sqrt{\frac{2}{\pi}}\frac{\sigma(\Delta t)}{\Delta t}.
\end{equation}

In a stationary equilibrium ensemble, the variance of the velocity increment $\sigma^2(\Delta t)$ 
is completely determined by the velocity autocorrelation function (VACF) as $\sigma^2(\Delta t) = 
2 \left[ 1- C_{vv}(\Delta t) \right]$ (see App. G). This relationship gives $S_v(\Delta t)$ 
in terms of the equilibrium VACF as
\begin{equation}
S_v(\Delta t) =\frac{2}{\Delta t} \sqrt{  \frac{1- C_{vv}(\Delta t)}{\pi}}.
\label{eq_sda}
\end{equation}
Since $1- C_{vv}(\Delta t)  \sim \Delta t^{2 H_v}$ one finds $S_v(\Delta t) \sim \Delta t^{H_v-1}$, and in the case that fluid-inertial effects are included (see App. G) recovers $H_v=1/4$.

Additionally, from the conditional covariance formula for multivariate Gaussian distributions, we find that in the non-equilibrium case in which the particle is conditioned on an initial velocity, the velocity increment variance is $\sigma^2(\Delta t) =  1- \left[C_{vv}(\Delta t) \right]^2$. This form recovers (see App. G.1) the same non-equilibrium fractal scaling law $H_v=1/4$, which is to state again that the equilibrium fluctuations govern the relaxation of unthermalized initial states.

\begin{figure}
 \includegraphics[width=8cm]{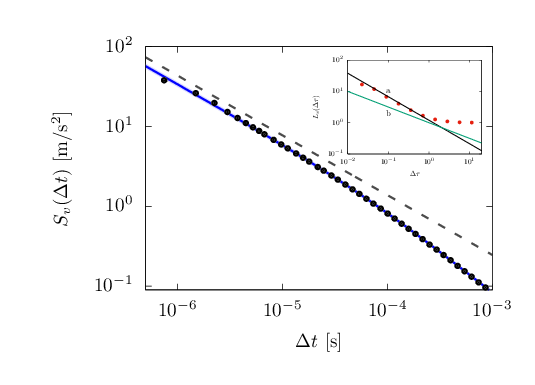}
    \caption{Hydrodynamic fractal scaling of Brownian velocity fluctuations. The black circles are $S_v(\Delta t)$ from experimental data of a barium titanate glass microsphere optically trapped in acetone after accounting for laser noise and finite experimental resolution as detailed in Appendix H. Statistical error bars \cite{flyvbjerg} are smaller than the marker size, and parametric uncertainty is detailed in Appendix H. The solid blue line is $S_v(\Delta t)$ as predicted by the hydrodynamic GLE. The grey dashed line is the asymptotic scaling $\Delta t^{-3/4}$. The inset shows the scaling of $L_u(\Delta \tau)$ vs $\tau$ using raw data without laser noise or finite-time corrections ($\bullet$). Line (a) represents the scaling $\Delta \tau^{-3/4}$, line (b) $\Delta \tau^{-1/2}$. }
    \label{FigSDA}
\end{figure}

We note that $L_u(\Delta \tau)$  and $S_u(\Delta \tau)$ asymptotically converge when the variance of the signal increments dominates the time step, yielding an expected squared slope that is much larger than unity (App. A). In figure \ref{FigSDA} we plot the theoretical curves for both $S_v(\Delta t)$ and $L_u(\Delta\tau)$ alongside the experimental data for our barium titanate glass microsphere in acetone. The $S_v(\Delta t)$ data plotted is found applying eq. (\ref{eq_sda}) on six independent equilibrium velocity traces of length 83.886 ms, and accounting for independent noise in the laser (App. H). The data is scaled to account for finite time resolution effects as discussed in Appendix H. $S_v(\Delta t)$
 recovers the short fractal signature, a consequence of the fluid inertial effects acting on Brownian hydromechanics, and thus establishes the non-equilibrium universality class of Brownian motion in fluid media possessing a finite non-vanishing density.

% ...
To analyze
further the dynamic origin of the 7/4-law for the velocity fractal dimension, we perform direct stochastic simulations of Brownian motion
including fluid inertial effects. To account for finite viscoelastic relaxation at the shortest timescales and enable numerical integration, we regularize the singular Basset kernel using a high-frequency cutoff parameter $\Gamma$ (where the limit $\Gamma \to \infty$ strictly recovers the singular behavior of the Basset kernel). The full Markovian embedding and Euler-Langevin integration scheme are detailed in Appendix B.

\begin{figure}
\includegraphics[width=8cm]{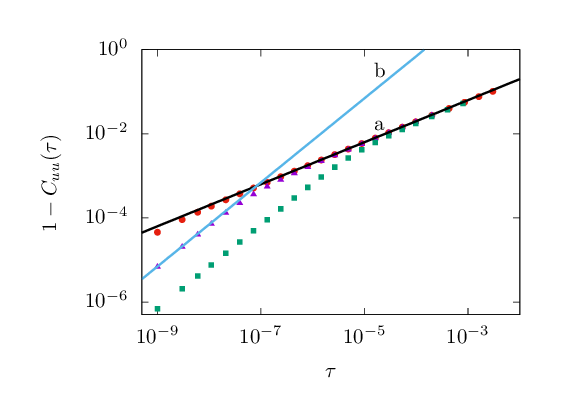}
\caption{Initial scaling of $1-C_{uu}^{(n)}(\tau)$ vs $\Delta \tau$ 
 of the solutions of the inertial
model eq. (\ref{eq9}) at  $\kappa_s=0$, $\beta=1$ for different values
of $\Gamma$. Symbols ($\bullet$) refer to $\Gamma=10^{10}$,
($\blacktriangle$) to $\Gamma=10^8$, ($\blacksquare$) to
$\Gamma=10^6$. Line (a) corresponds to
$1-C_{uu}^{(n)}(\tau)\sim \tau^{1/2}$, line (b) to
$1-C_{uu}^{(n)}(\tau)\sim \tau$.}
\label{Fig4}
\end{figure}

The variable regularity analysis (varying $\Gamma$) confirms the statement that the inertial Basset kernel modifies the universal non-equilibrium properties of Brownian dynamics with respect to the Einstein-Langevin  theory.
Figure \ref{Fig4} depicts the short-term scaling
of the normalized velocity autocorrelation function $1-C_{uu}(\tau)$
obtained from the hydrodynamic stochastic simulation at different values of $\Gamma$.
A crossover from  the $1-C_{uu}^{(n)}(\tau)\sim \tau$ to  
the $1-C_{uu}^{(n)}(\tau)\sim \tau^{1/2}$ scaling occurs at ``small''
values of $\Gamma$ due to the boundedness of $k(\tau=0)$.
At $\Gamma=10^{10}$ the $\tau^{1/2}$-scaling occurs for
$\tau >10^{-8}$, i.e.  eight order of magnitude smaller than
the momentum relaxation time. While physical liquids are constrained by finite viscoelastic limits, setting $\Gamma$ to $10^{10}$ provides a numerical benchmark to isolate the singular Basset limit and verify the scaling laws of the fluid-inertial universality class.

We can therefore summarize the different hydromechanic
regimes in terms of the short-time scaling of
the velocity autocorrelation function,
\begin{equation}
1-C_{uu}(\tau) \sim
\left \{
\begin{array}{lll}
\tau^{1/2} & \;\;\;\; & \mbox{fluid-inertial effect} \\
\tau & &  \mbox{Einstein-Langevin} \\
\tau^2 & & \mbox{Lipschitz-continuous}
\end{array}
\right .
\label{eq_app3}
\end{equation}
for small $\tau$'s,
that can be generalized in the form $1-C_{uu}(\tau) \sim \tau^{\beta}$ with
$\beta= 2 \, H_v$.

To conclude, short-time measurements
have unlocked access to the non-equilibrium properties of Brownian motion. As shown in this Letter,
and addressed in \cite{btpg}, these non-equilibrium phenomena 
gravitate around the regularity properties of  velocity fluctuations.
The linchpin of the theory is the velocity H\"older exponent $H_v$
or, equivalently, its fractal dimension $d_v$.
The value $d_v=7/4$ of  the velocity 
fractal dimension in the presence of inertial
effects is a consequence of the fluid inertia modulating the correlation properties of
the thermal-hydrodynamic force $R(t)$, that is
the distributional derivative of a stochastic process
possessing H\"older exponent $1/4$. This follows
also from the free-inertial process formalism developed
in \cite{btpg}. 

Fundamentally, equations for the dynamic scaling show that
field interactions modify the universality class of Brownian
motion dynamics, and this is a non-equilibrium effect, i.e.
it involves short-time dynamics.
This result and the approach adopted here could be extended to any system-bath interaction characterized by memory effects. The relation between the dynamic scaling of exponents and the singularity of a generalized fractional memory kernel is analyzed in Appendix C.

In this regard, the short-time scaling of the velocity
autocorrelation function  plays a central role, as the
behavior of $1- C_{uu}(\tau)$ depends on the
regularity of $u(\tau)$ and ultimately on the
fluid-dynamic interactions controlling the values of $H_v$ or $d_v$.

\newpage
\onecolumngrid

\section*{Appendices}

\twocolumngrid

\section{A. Notes on the scale-dependent absolute slope}

It is important to specify the key differences in the scale-dependent absolute slope and the length resolution analysis, each intending to resolve fractal scalings. For any general curve $c(\tau)$, $L_c(\Delta \tau)$ yields the proper geometric length and true fractal dimension, while $S_c(\Delta \tau)$ may convey more information about the signal's asymptotic fluctuations. For example, $L_c(\Delta \tau)$ for a stationary signal $c(\tau)$ in the large-scale limit will saturate at 1, while $S_c(\Delta \tau)$ will decay as $\Delta \tau^{-1}$. Further, as noted in Appendix G, the two may differ in the small-scale limit for Lipschitz-continuous signals $c(\tau)$, in which case  $L_c(\Delta \tau)$ always gives unity but $S_c(\Delta \tau)$ does not. As stated in the main text, $L_c(\Delta \tau)$  and $S_c(\Delta \tau)$ agree when the variance of the signal increments dominates the time step.

\section{B. Markovian embedding and Euler-Langevin integration scheme}
We expand here on the Markovian Embedding of the Hydrodynamic GLE and the Euler-Langevin integration scheme. One can approximate the fluid inertial kernel $k(t)$
as $k(\tau)=b \, g(\tau,\Gamma,N_i)= \sum_{k=1}^{N_i}\gamma_k e^{-\mu_k \, \tau}$ where 
\begin{equation}
g(\tau,\Gamma,N_i)=  \sqrt{\Gamma} \sum_{k=1}^{N_i} \frac{1}{a^{k/2}}
e^{-t \, \Gamma/a^k} = \sum_{k=1}^{N_i} \overline{\gamma}_k  \, e^{-\mu_k \, \tau}
\label{eq8}
\end{equation}
with $a=6$, $N_i=20$, where $\Gamma>0$, and   $\gamma_k= b \, \overline{\gamma}_k$ (see App. F).
For $\tau > \tau_{\rm min}$ and $\tau \leq \tau_{\rm max}$, where
$\tau_{\rm min}=O(\Gamma^{-1})$ and $\tau_{\rm max} =O(10^4)$,
$g(\tau \Gamma, N_i) \simeq 1/\sqrt{\tau}$, and the differences between
the two functions are practically  indistinguishable (see App. F).
 For $\tau \rightarrow 0$,
 $g(\tau,\Gamma,N_i)$ attains a finite value  $g(0;\Gamma,N_i) 
\sim \sqrt{\Gamma}$,
and this phenomenology matches expectations for a physically realizable memory kernel for the shortest timescales as it accounts for the viscoelastic effects 
on the fluid-inertial
kernel \cite{procopiogiona}.  The upper bound  $\tau_{\rm max}$ stems from the
choice of a reasonably small number of exponential
modes $N_i=20$, while $\tau_{\rm max} \rightarrow \infty$ for $N_i \rightarrow \infty$.
Similarly, one can strictly recover the singular behavior
of the Basset kernel at small $\tau$, simply considering the limit $\Gamma
\rightarrow \infty$.

The decomposition of the kernel $k(\tau)$ in terms
of exponentially decaying modes provides a simple and general
way to account for the inertial effects within a Markovian
embedding of the dynamics consistent with the Kubo  second 
fluctuation-dissipation
relation (for details see \cite{gpp1}).
By introducing a system of auxiliary memory variables 
$\{z_k(\tau)\}_{k=1}^{N_i}$,
each of which associated with a specific mode of the 
kernel expansion eq. (\ref{eq8}),  the Brownian dynamics in a trap attains the form

\begin{align} 
\frac{d x(\tau)}{d \tau} &= u(\tau) \nonumber \\[1ex] 
\begin{split}
\frac{ d u(\tau)}{d \tau} &= - u(\tau) - \kappa_s \, x(\tau)- G \, u(\tau) + \\ 
&\quad \sum_{k=1}^{N_i} \gamma_k \, \mu_k \, z_k(\tau) + \sqrt{2} \, \sum_{k=1}^{N_i} d_k \, \xi_k(\tau) + \sqrt{2} \, \xi(\tau) 
\end{split} \label{eq9} \\[1ex] 
\frac{d z_k( \tau)}{d \tau} &= - \mu_k \, z_k(\tau) + u(\tau) + \sqrt{2} \, c_k\, \xi_k(\tau) \nonumber 
\end{align}
where  $\xi(\tau)$ and $\xi_k(\tau)$, $k=1,\dots,N_i$
are distributional derivatives of independent Wiener processes,
$G=\sum_{k=1}^{N_i} \gamma_k$, and
the coefficients $d_k$, $c_k$  are given by
$d_k= \sqrt{\gamma_k}$, $c_k= - 1/\sqrt{\gamma_k}$, $k=1,\dots,N_i$
\cite{gpp1},
in order to fulfill the Kubo fluctuation-dissipation relations of the
first and second kind \cite{kubo2}.
These equations can be numerically integrated using a Euler-Langevin algorithm
with a time-step $h_\tau =O(10^{-2}/\sqrt{\Gamma})$, in order to
ensure stability of the algorithm, i.e. $h_\tau \mu_1 < 10^{-2}$.

\begin{figure}
\includegraphics[width=8cm]{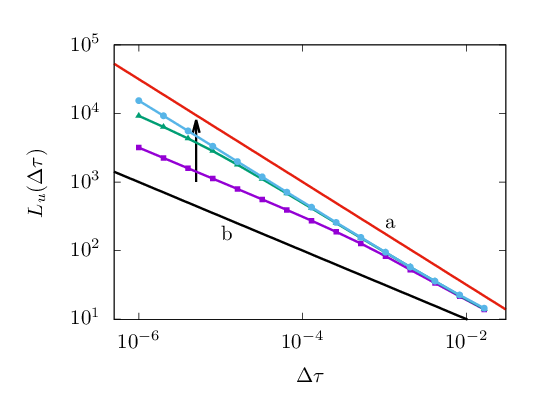}
\caption{Length-resolution analysis $L_u(\Delta \tau)$ vs $\Delta \tau$ of the
velocity realizations  solution of the inertial
model eq. (\ref{eq9}) at  $\kappa_s=0.1$, $b=0.1$ for different values
of $\Gamma$. The arrow indicates increasing values
of $\Gamma=10^4,\,10^6,\,10^8$. Line (a)  represents the scaling
$L_u(\Delta \tau) \sim \Delta \tau^{-3/4}$,
line (b) the scaling $L_u(\Delta \tau) \sim \Delta \tau^{-1/2}$.}

\label{Fig3}
\end{figure}

Figure \ref{Fig3} depicts the results of the length-resolution analysis of this
process for different values of $\Gamma$. 
For small values of $\Gamma=10^4, \, 10^6$, a crossover occurs
between the $3/4$-scaling and the $1/2$-scaling at resolutions order of $\Gamma^{-1}$.
This stems from the regularity  at $t=0$ of the kernels considered.
At $\Gamma=10^8$,  the $3/4$-law is the
only  observed scaling for  $\Delta \tau>10^{-6}$
as, in this range of temporal resolution,
the saturation of $k(\tau)$ occurring at smaller time scales is immaterial 
as regards particle dynamics. 

\section{C. Generalized Fractional Memory Kernels}

\begin{figure}
\includegraphics[width=8cm]{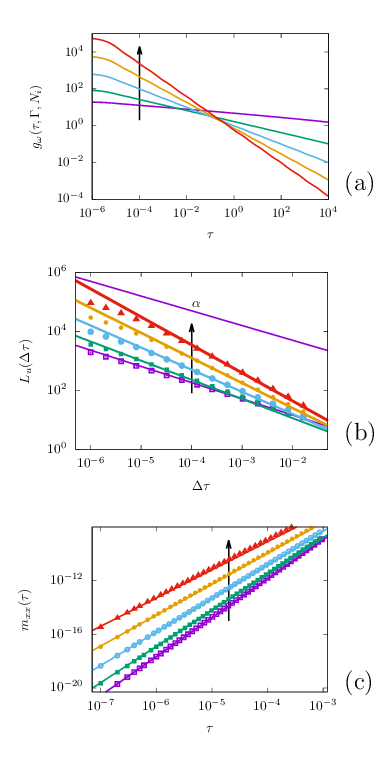}
\caption{Panel (a) : Function  $g_\omega(\tau,\Gamma,N_i)$ vs $\tau$ 
for $N_i=20$ at 
$\Gamma=10^6$ for increasing values of $\omega$.
Panel (b) : Length-resolution analysis,  $L_u(\Delta \tau)$ vs 
$\Delta \tau$ (symbols)
  of a realization of the particle
velocity deriving from the solution of eq. (\ref{eq9}) in the presence of the
inertial kernel $k(\tau)= b \, g_\omega(\tau,\Gamma,N_i)$, $a=6$, $b=\kappa_s=0.1$
 where $g_\omega(\tau,\Gamma,N_i)$
 is depicted in panel (a) for increasing values of $\omega$.
Solid lines represents the scaling $L_u(\Delta \tau) 
\sim \Delta \tau^{-(1+\omega)/2}$. Line ($\alpha$) represents
the Wiener scaling $L_u(\Delta \tau) \sim \Delta \tau^{-1/2}$.
Panel (c) : $m_{xx}(\tau)$ vs $\tau$  in the presence of the
inertial kernel $k(\tau)= b \, g_\omega(\tau,\Gamma,N_i)$,
$\beta=\kappa_s=0.1$,
for increasing values of $\omega$. Symbols
represent the simulation results from the moment dynamics,
solid lines correspond to the scaling laws $m_{xx}(\tau) \sim 
\tau^{2+(1-\omega)}$.
The arrows indicate increasing values of $\omega=0.1,\,0.3,\,0.5,\,0.7,\,0.9$.}
\label{Fig5}
\end{figure}

In order to unveil the relation between the singularity
of the inertial kernel and the universality class in non-equilibrium conditions,
it is useful to analyze  a more general
  family of  kernels $k(\tau) \sim \tau^{-\omega}$, 
$ \omega \in (0,1)$ parametrically dependent on the singularity exponent
$\omega$ at $\tau=0$, figure \ref{Fig5} panel (a).
This can be achieved 
 by setting 
$k(\tau)= b \, g_\omega(\tau,\Gamma,N_i)$, where
\begin{equation}
g_\omega(\tau,\Gamma,N_i)=   \sum_{k=1}^{N_i}  \left ( \frac{\Gamma}{a^{k}}
\right )^\omega
e^{-\tau \, \Gamma/a^k} 
\label{eq11}
\end{equation} 
With respect to eq. (\ref{eq8}) the rates $\mu_k$ do not change,
while $\gamma_k= b \, (\Gamma/a^k)^\omega$.
The stochastic model eq. (\ref{eq9})  with the
$\gamma_k$'s and $\mu_k$'s defined by eq. (\ref{eq11})
satisfies  the Kubo fluctuation dissipation
relations,  as $d_k=\sqrt{\gamma_k}$ and $c_k=-1/\sqrt{\gamma_k}$,
$k=1,\dots,N_i$.

The hydrodynamic case (Basset inertial kernel) corresponds to $\omega=1/2$.
Figure \ref{Fig5} panel (a) depicts the scaling of these  kernels
for  different values of $\omega$. The analysis grounded
on eq. (\ref{eq8}) can be applied also to this
case. The results for $m_{xx}(\tau)$ (starting from the initial
condition $x(0)=u(0)=0$), and for $L_u(\Delta \tau)$ are depicted
in panels (b) and (c) of figure \ref{Fig5}. Also in this
case eq. (\ref{eq4}) applies, where the exponent $\varphi$ is given
by
\begin{equation}
\varphi=3-\omega
\label{eq12}
\end{equation}
This completes the analysis of the relations between the 
structure of the kernel $k(\tau)$ and the dynamic Brownian motion
exponents $\varphi$, $H_v$ and $d_v$.

\section{D. Experimental system}
\label{smsec1}
A detailed description of the experimental apparatus and data processing for this study can be found in \cite{btr}. A barium titanate glass microsphere (with density $4200\pm 200$ kg/m$^3$ and radius $3.4\pm0.1 \mu$m) is optically trapped in acetone (with density $790$ kg/m$^3$ and viscosity $0.32$ mPas), and its position is measured using a cut mirror and split beam detection \cite{hres4} on a custom built high-power balanced photodetector. The power difference incident on the two ports of the balanced photodetector changes linearly with the single axis displacement of the microsphere within the range of the trap. The output voltage is measured with a 16-bit digitization card at 200 Msamples/s for $2^{24}$ samples. We account for the detector's built in high pass filter by measuring its response function and inverting its effects on the data with Tikhonov regularization scheme. We use a bin number of 150, set by laser noise, to achieve a temporal resolution of 750ns. We fit the equilibrium MSD to the hydrodynamic theory up to 1.2 ms, accounting for the added mass, to extract the particle's radius, the trap strength, and the volts-to-meter conversion factor. We then apply an 8th-order finite differencing scheme to the position traces to obtain the velocity traces that we analyze in this work. The resulting velocity distribution has a standard deviation $98\%$ the Maxwell-Boltzmann distribution with an SNR of 12 dB.

\section{E. Universality classes}
\label{smsec2}
In statistical physics, universality is meant to indicate
properties of systems that do not depend on the fine
details of the interactions, topology, etc., but exclusively
on the physical dimension $d$ of the space in which the
interactions occur \cite{kadanoff}. A typical example
are the scaling properties of percolation models near the
critical threshold \cite{stauffer}.

The concept of universality has been progressively enlarged
to indicate  classes of ``qualitative equivalence'' of models
that share, in their dynamic evolution, some common
scaling. This is the case of the widespread
use of {\em KPZ universality} \cite{kpz1,kpz2},
to indicate growth models 
characterized by long time behaviors ``similar to that of the KPZ 
equation itself'' \cite{kpz3}.
In this broad meaning, we use the concept of universality class
for physical Brownian motion, to indicate the common
scaling properties of Brownian motion characterized
by specific  hydromechanic interactions with a solvent
fluid at thermal equilibrium.

\section{F. Fluid-particle hydromechanics}
\label{smsec3}

Let $\widehat{v}(s)$ be the Laplace transform of $v(t)$,
and analogously for the other quantities.
Consider the time-dependent Stokes equation for the
hydrodynamics of a Newtonian incompressible liquid,
the Laplace transform $\widehat{F}_{f \rightarrow p}[\widehat{v}(s)]$
of the force exerted by the fluid onto a spherical rigid
particle of radius $R_p$ is given by \cite{kim}

\begin{multline}
\widehat{F}_{f \rightarrow p}[\widehat{v}(s)]= - 6 \, \pi \mu \, R_p \,
\widehat{v}(s) - \\
6 \, \pi \, \sqrt{ \mu \, \rho} \, R_p^2 \,
\frac{1}{\sqrt{s}} (s \, \widehat{v}(s) ) - \frac{2}{3} \pi\, R_p^3 \, (s \, \widehat{v}(s) )
\label{eqh1}
\end{multline}
In the present analysis we neglect the last term, associated
with the added-mass effect, as it vanishes
once compressibility effects  are taken into account.

As regards the second term, indicating with ${\mathcal L}^{-1}$
the inverse Laplace transform (and ${\mathcal L}$ the direct one), we have
\begin{equation}
{\mathcal L}^{-1} \left [ s \,  \widehat{v}(s) \right ]
= \frac{d v(t)}{d t}+ v(0)\, \delta(t)
\label{eqh2}
\end{equation}
motivating the occurrence of the term $u(0) \, \delta(\theta)$
in the non-dimensional equations for the particle
dynamics, eq. (\ref{eq2}).

Next, consider the nondimensional eq. (\ref{eq2}) in the main text.
If the value of $k(\tau=0)$ is bounded then
\begin{multline}
    \int_0^\tau k(\tau-\theta) \, \left ( \frac{d u(\theta)}{d \theta}
+ u(0) \, \delta(\theta) \right )  \, d \theta= \\
 k(0) \, u(\tau) - \int_0^\tau \frac{d k(\tau-\theta)}{d \theta} \, u(\theta)
\, d \theta
\label{eqh3}
\end{multline}
If $k(\tau)$ is not bounded it is not possible to
express the stochastic forcing in terms of elementary stochastic
processes \cite{gpp1}.
In fact, the condition of bounded $k(0)$ admits a physical
origin, as it is related to the finite velocity of
propagation of shear stress \cite{procopiogiona}.

\begin{figure}
\includegraphics[width=8cm]{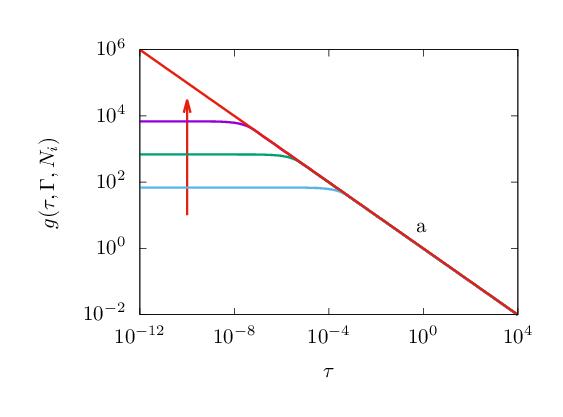}
\caption{Function $g(\tau,\Gamma,N_i)$ vs $t$ defined by eq. (\ref{eqh4})
for  $a=6$ and $N_i=20$ at different values
of $\Gamma$. The arrow indicates increasing values of $\Gamma=10^4,\,10^6,\,10^8$. Line (a) represents the function $1/\sqrt{\tau}$.}
\label{sFig1}
\end{figure}

In the direct simulations of Brownian dynamics including
fluid inertial effects,
the fluid-inertial kernel $k(\tau)$ is expressed 
 as  $k(\tau)=b \, g(\tau,\Gamma,N_i)$, where
\begin{equation}
g(\tau,\Gamma,N_i)=  \sqrt{\Gamma} \sum_{k=1}^{N_i} \frac{1}{a^{k/2}}
e^{-\tau \, \Gamma/a^k} = \sum_{k=1}^{N_i} \overline{\gamma}_k  \, e^{-\mu_k \, \tau}
\label{eqh4}
\end{equation}
with $a=6$, considering $N_i=20$ modes.
Figure \ref{sFig1} depicts the function $g(\tau,\Gamma,N_i)$ 
truncated to $N_i=20$, for different values of the parameter $\Gamma$,
which shows that qualitative properties of this function discussed
in the main text.

\section{G. Short-time scaling and correlations}
\label{smsec4}

Consider the quantity $\langle (v(t+\Delta t)- v(t))^2 \rangle_{\rm eq}$
evaluated with respect to the probability measure of equilibrium
fluctuations.
Since 
\begin{equation}
\langle  |v(t+\Delta t)- v(t) | \rangle_{\rm eq} \sim \Delta t^{H_v}
\label{eqs1}
\end{equation}
at short $\Delta t$, defines the H\"older exponent
of the particle velocity, it follows that
\begin{equation}
\langle (v(t+\Delta t)- v(t))^2 \rangle_{\rm eq} \sim \Delta t^{2 \, H_v}
\label{eqs2}
\end{equation}
But,
\begin{multline}
\langle (v(t+\Delta t)- v(t))^2 \rangle_{\rm eq}
= 2 \, \langle v^2 \rangle_{\rm eq} - 2 \,
\langle v(t+\Delta t) \, v(t) \rangle_{\rm eq}
 \\
= 2 \, C_{vv}(0) \, \left [1- C_{vv}^{(n)}(\Delta t) \right ]
\label{eqs3}
\end{multline}
where $C_{vv}^{(n)}(t)= C_{vv}(t)/C_{vv}(0)$ is the
normalized velocity autocorrelation function in equilibrium
conditions.
Eqs. (\ref{eqs2})-(\ref{eqs3}) indicate that the  H\"older exponent $H_v$
is related to the short-time scaling of the velocity
autocorrelation function at equilibrium.

The short-time scaling of $C_{vv}^{(n)}(t)$  follows directly
from  the particle dynamics (eq. (\ref{eq2}) in the main text) that
in dimensional form, neglecting the harmonic potential term
(inessential in the short-time analysis), reads
\begin{multline}
m \, \frac{d v(t)}{d t} = - \eta \, v(t) -\\
\int_0^t
k(t-t^\prime) \left ( \frac{d v(t^\prime)}{d t^\prime} + v(0) \, 
\delta(t^\prime) \,  \right )  \, d t^\prime + R(t)
\label{eqs4}
\end{multline}
The stochastic forcing $R(t)$ satisfies Kubo fluctuation
dissipation relations, and thus the Langevin
condition $\langle R(t) v(0) \rangle_{\rm eq}=0$ for $t \geq 0$.
This implies that the normalized velocity autocorrelation
function $C_{vv}^{(n)}(t)$ satisfies the equation
\begin{multline}
m \, \frac{d C_{vv}^{(n)}(t)}{d t} = - \eta \, C_{vv}^{(n)}(t) -\\
\int_0^t
k(t-t^\prime) \left ( \frac{d C_{vv}^{(n)}(t^\prime)}{d t^\prime} + 
\delta(t^\prime)   \right )  \, d t^\prime
\label{eqs5}
\end{multline}
with $C_{vv}^{(n)}(0)=1$.
Thus, its Laplace transform $\widehat{C}_{vv}^{(n)}(s)$ attains
the expression
\begin{equation}
\widehat{C}_{vv}^{(n)}(s)= \frac{m}{m \,s + \eta + \widehat{k}(s)}
\label{eqs6}
\end{equation}
so that
\begin{multline}
{\mathcal L} \left [1- C_{vv}^{(n)}(t) \right ] = \frac{1}{s} - \frac{m}{m \,s + \eta + \widehat{k}(s)}
\\
= \frac{\eta+ \widehat{k}(s)}{s\, (m \, s + \eta + \widehat{k}(s))}
\label{eqs7}
\end{multline}
The short time scaling of $1- C_{vv}^{(n)}(t)$ corresponds to
the scaling of $ s \, {\mathcal L} \left [1- C_{vv}^{(n)}(t) \right ]$
for $s  \rightarrow \infty$. Considering
that for large $s$, $\widehat{k}(s) \sim s^{-1/2}$,
 the
leading-order term at the numerator of eq. (\ref{eqs7}) is $\widehat{k}(s)$
while the denominator is controlled by $m \, s^2$. Thus
\begin{equation}
{\mathcal L} \left [1- C_{vv}^{(n)}(t) \right ] \simeq 
\frac{\widehat{k}(s)}{m \, s^2} \sim \frac{1}{s^{3/2}}
\label{eqs8}
\end{equation}
and therefore, at short time scales
\begin{equation}
1- C_{vv}^{(n)}(t)  \sim t^{1/2}
\label{eqs9}
\end{equation}
that implies 
\begin{equation}
\langle (v(t+\Delta t)- v(t))^2 \rangle_{\rm eq} \sim \Delta t^{1/2}
\label{eqs10}
\end{equation}
i.e. $H_v=1/4$ and $d_v=7/4$. 

Next, consider the Einstein-Langevin model (i.e. $k(t)=0)$.
In this case, for large $s$ we have
\begin{equation}
{\mathcal L} \left [1- C_{vv}^{(n)}(t) \right ]=
\frac{\eta}{s \, (m \, s+ \eta)} \simeq \frac{\eta}{m\, s^2}
\label{eqs11}
\end{equation}
that corresponds in the time-domain to $1- C_{vv}^{(n)}(t) \sim t$.

As a model possessing Lipschitz-continuous velocity realizations,
one can consider the case of Brownian motion in a Maxwell fluid in
the absence of fluid-inertial effects.
Particle dynamics attains in this case the form
\begin{equation}
m \, \frac{d v(t)}{d t }=- \eta \, \lambda \,e^{-\lambda \, t} * v(t) +
R(t)
\label{eqs12}
\end{equation}
where  ``$*$'' indicates convolution, 
$\lambda>0$ is the relaxation time, and $R(t)$ is
such to satisfy the Kubo fluctuation-dissipation relations.
This model admits the modal representation
\begin{eqnarray}
m \, \frac{d v(t)}{d t} &= & - \eta \, \lambda \, \theta(t) \nonumber \\
\frac{d \theta(t)}{d t} & =  & - \lambda \, \theta(t) + v(t) +\sqrt{2}
\, {\tilde b} \, \xi(t)
\label{eqs13}
\end{eqnarray}
where $\xi(t)= d w(t)/d t$ is the distributional
derivative of a Wiener process $w(t)$, and ${\tilde b}= \sqrt{k_B \, T/\eta}$.
Performing the same calculations as above, one obtains for
the normalized velocity autocorrelation function
\begin{equation}
{\mathcal L} \left [1- C_{vv}^{(n)}(t) \right ]=
\frac{\eta \, \lambda}{ m \, s^2 \, (s+ \lambda)+ \eta \, \lambda \, s}
\label{eqs14}
\end{equation}
Therefore for large $s$, 
${\mathcal L} \left [1- C_{vv}^{(n)}(t) \right ] \sim 1/s^3$ implying
\begin{equation}
1- C_{vv}^{(n)}(t) \sim t^2
\label{eqs15}
\end{equation}
Gathering these scaling results, eq. (\ref{eq_app3}) in the main text follows.

Next, consider the probabilistic interpretation of
the length-resolution analysis,
\begin{equation}
L_v(\Delta t) = \frac{1}{\Delta t} \, \left \langle 
\sqrt{\Delta t^2 + \Delta v^2(\Delta t)} \right \rangle_{\rm eq}
\label{eqs16}
\end{equation}
where $\Delta v(\Delta t)= v(t+\Delta t)- v(t)$.
The probability density function $p_{\Delta v, {\rm eq}}(\Delta v ; \Delta t)$ 
for the variable  $\Delta v(\Delta t)$ is Gaussian, possessing zero mean 
and squared variance $\sigma_{\Delta v}^2(\Delta t)$ given by
\begin{equation}
\sigma_{\Delta v}^2(\Delta t)= 2 \, C_{vv}(0) \, \left [
1- C_{vv}^{(n)}(\Delta t) \right ]
\label{eqs17}
\end{equation}
Thus,
\begin{equation}
L_v(\Delta t) = \frac{1}{ \Delta t} \int_{-\infty}^\infty \sqrt{\Delta t^2 +
\sigma_{\Delta v}^2(\Delta t) \, y^2} \, \frac{e^{-y^2/2}}{\sqrt{2 \, \pi}}
\, d y,
\label{eqs19}
\end{equation}
using the change of variables $y=z/\sigma_{\Delta v}(\Delta t)$. At short $\Delta t$, $\sigma_{\Delta v}^2(\Delta t)$ 
dominates with respect to $\Delta t^2$ so that
\begin{multline}
L_v(\Delta t) \simeq \frac{\sigma_{\Delta v}(\Delta t)}{\Delta t}
\, \int_{-\infty}^\infty |y| \, \frac{e^{-y^2/2}}{\sqrt{2 \, \pi}}
\, d y \sim \frac{\sigma_{\Delta v}(\Delta t)}{\Delta t}\\
\sim \frac{\Delta t^{1/4}}{\Delta t} = \Delta t^{-3/4}
\label{eqs20}
\end{multline}
corresponding to a fractal dimension $d_v=7/4$.

\subsection{G.1 The non-equilibrium case}

The covariance formula for conditional multivariate gaussians is 

\begin{equation}
    \sigma^2_{XX|(Y=y)} = \sigma^2_{XX} - \frac{\sigma_{XY}^2}{\sigma_{YY}^2}.
    \label{multivariategauss}
\end{equation}
In the case that we are considering the velocity signal, we recognize the constant velocity variance $\sigma_{XX}^2 = \sigma^2_{YY} = k_BT/m$, which is non-dimensionalized to unity. This allows us to write the conditional variance of the velocity increment in terms of the equilibrium VACF as
\begin{equation}
\sigma_{vv, {\rm neq}}^2(\Delta \tau) \sim  1- \left[C_{vv}^{(n)}(\Delta \tau) \right]^2.
\end{equation}
Throughout these appendices we use $\tau$  and $\Delta \tau$ both as a dimensional
(as in the present case)
or nondimensional time variables as their meaning follows directly from the context.

For the hydrodynamic case, we substitute into this equation the form $C_{vv}^{(n)}(\Delta \tau)\sim1-A \, \Delta \tau^{1/2}$ from eq. (\ref{eqs9}), where A is some constant prefactor. We find the same leading short time scaling behavior as in the case of the equilibrium velocity trace, and thus the same fractal dimension

\begin{multline}
    \sigma_{vv, {\rm neq}}^2(\Delta \tau) \sim 1-(1-A\, \Delta \tau^{1/2})^2 =\\
     2 \, A \, \Delta \tau^{1/2}-A^2 \Delta \tau \sim  \Delta \tau^{1/2}.
\end{multline}

\begin{figure}
\includegraphics[width=8cm]{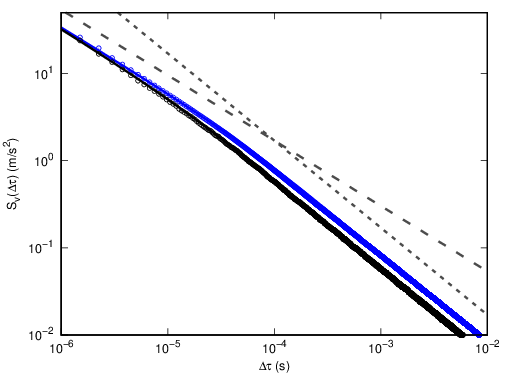}
\caption{The hydrodynamic $S_v(\Delta \tau)$ obtained from equilibrium, and non-equilibrium velocity traces of a barium titanate glass microsphere in acetone. The solid blue line and solid black lines represent the theoretical curves for $S_v(\Delta \tau)$ predicted by the hydrodynamic GLE for the equilibrium and non-equilibrium (starting from $v(0)\approx0$) cases respectively. The blue circles are from equilibrium traces, and the black circles are data from conditioned non-equilibrium traces. Statistical error bars are strictly smaller than the markers and are omitted. }
\label{sda_neq}
\end{figure}

Doing the same calculation for non-equilibrium velocity traces in air (Einstein-Langevin), we find the lowest order term yields $\sigma_{vv, {\rm neq}}^2(\Delta \tau) \sim  \Delta \tau$, also agreeing with the equilibrium case. This is a statement of the Onsager regression hypothesis, stating that the equilibrium fluctuations govern the relaxation of unthermalized initial states. This theoretical convergence is empirically validated in Figure \ref{sda_neq}, which plots the analytic curves and experimental data for $S_v(\Delta \tau)$. As predicted, both the equilibrium and non-equilibrium velocity time traces have the same asymptotic scaling laws.

Now we turn to consider the position time-traces and the corresponding scale-dependent absolute slope 
$S_x(\Delta \tau)$ which obtains units of velocity. We note that the equilibrium displacement variance 
$\sigma_{xx}^2(\Delta \tau)$ is simply the Mean Squared Displacement (MSD), which gives the SDV as

\begin{align}
S_{x}(\Delta \tau) &= \frac{1}{\Delta \tau} \sqrt{\frac{2}{\pi} \text{MSD}(\Delta \tau)}.
\label{eq_sdv_}
\end{align}

\begin{figure}
\includegraphics[width=8cm]{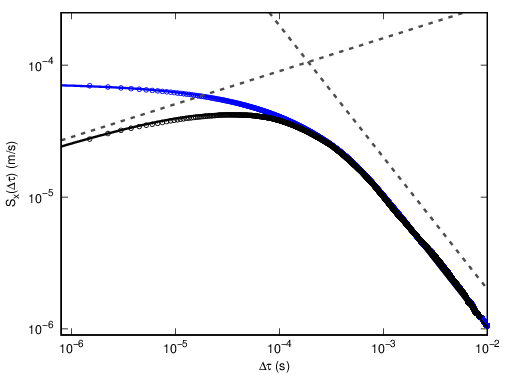}
\caption{The hydrodynamic $S_x(\Delta \tau)$ obtained from equilibrium, and non-equilibrium position traces of a barium titanate glass microsphere in acetone. The solid blue line and solid black lines represent the theoretical curves for $S_x(\Delta \tau)$ predicted by the hydrodynamic GLE for the equilibrium and non-equilibrium (starting from $v(0)\approx0$) cases respectively. The blue circles are from equilibrium traces, and the black circles are data from conditioned non-equilibrium traces. Statistical error bars are strictly smaller than the markers and are omitted.}
\label{sdv_neq}
\end{figure}

One may apply an equilibrium or non equilibrium MSD in eq. \ref{eq_sdv_} where the two are also related by eq. \ref{multivariategauss} as

\begin{align}
\text{MSD}_{\rm neq}(\Delta \tau)= \text{MSD}_{\rm eq}(\Delta \tau) - \left( \int_0^{\Delta \tau} C_{vv}(t) dt \right)^2 .
\label{neq_sdv}
\end{align}
when analyzing short-time non-equilibrium Brownian position traces as in \cite{btr, duplat} with superballistic MSD scalings laws $\varphi>2$, one finds $ L_{x}(\Delta \tau \rightarrow0) = \sqrt{\Delta\tau^2 +\Delta\tau^{\varphi}}/\Delta\tau\rightarrow1 $ for the position trace length resolution, whereas $S_x(\Delta \tau)$ yields $\sqrt{\Delta\tau^{\varphi}}/\Delta\tau=\Delta\tau^{(\varphi/2)-1}$, unmasking the superballistic behavior. Explicitly, we have the positive scalings $S(\Delta \tau) \sim \Delta\tau^{1/2}$, and $S(\Delta \tau) \sim \Delta\tau^{1/4}$ which correspond to the superballistic $\varphi=3$, and $\varphi=5/2$ for the Einstein-Langevin and Basset-Boussinesq cases respectively. Both scalings lead to $H_x>1$, which would imply that $x(t)$ is constant, and thus the scaling of $S_x(\Delta \tau)$ under these conditions fails to predict the true H\"older exponent $H_x=1$. The distinct behavior of $S_x(\Delta \tau)$ under these conditions is illustrated in figure \ref{sdv_neq}. Notably, the velocity-conditioned non-equilibrium SDV exhibits a non-zero short-time scaling of $1/4$, agreeing with the superballistic $5/2$ mean squared displacement result, whereas the equilibrium SDV settles to a zero slope.

\begin{figure}
\includegraphics[width=8cm]{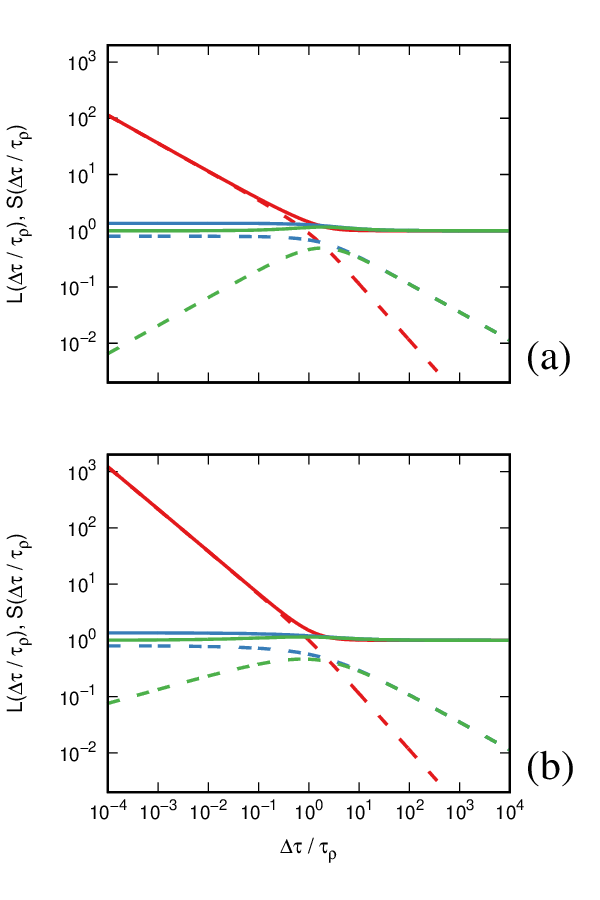}

\caption{Comparison of of $L_x(\Delta \tau)$, $S_x(\Delta \tau)$, $S_{x, neq}(\Delta \tau)$ (the non-equilibrium position trace scale-dependent slope), $L_v(\Delta \tau)$, and $S_v(\Delta \tau)$ in the cases of the a) Einstein-Langevin and b) Basset Boussinesq formulations of Brownian motion respectively. In the top panel a) the red solid line represents $L_v(\Delta \tau)$, the red dashed line represents $S_v(\Delta \tau)$, the green solid line represents $L_{x, neq}(\Delta \tau)$, the green dashed line $S_{x, neq}(\Delta \tau)$, the blue solid line $L_{x}$, and the blue dashed line $S_{x}$. In the bottom panel b) the curves are the same but for the Basset-Boussinesq case.}
\label{sdv_sda_lr}
\end{figure}

Figure \ref{sdv_sda_lr} provides a comprehensive comparison of $S_v(\Delta \tau)$, $L_v(\Delta \tau)$, $S_x(\Delta \tau)$, and $L_x(\Delta \tau)$ for both the Einstein-Langevin and Basset-Boussinesq formulations. This figure graphically reinforces the divergence between true geometric fractal dimensions ($d_{v}=3/2$ in the Einstein-Langevin case versus $d_{v}=7/4$ in Basset-Boussinesq case) and the scaling exponents extracted from the scale-dependent absolute slope analysis. Note how $L_v(\Delta \tau)$ (red solid) and $S_v(\Delta \tau)$ (red dashed) agree below the momentum relaxation time $\tau_\rho$ displaying the fractal dimension of either the Einstein-Langevin or Basset-Boussinesq case. Past $\tau_\rho$ $L_v(\Delta \tau)$ saturates while $S_v(\Delta \tau)$ does not. Also note how $S_x(\Delta \tau)$ (blue dashed) handles the transition between the ballistic regime and diffusive regime scaling while $L_x(\Delta \tau)$ cannot.

\subsection{G.2 Effects of fluid density $\rho$  and crossovers} 
\label{eqsec4a}

The fluid inertial kernel $k(t)$ is proportional to the
square root of the fluid density. This implies in nondimensional
terms that, if $k(\tau) = b \, g(\tau,\Gamma,N_i)$ is
characterized in liquids by a prefactor $b$ order of $10^{-1}$-$10^0$,
the prefactor $b$ in gases is between one or two order of magnitude
smaller in normal conditions of temperature and pressure.

\begin{figure}
\includegraphics[width=8cm]{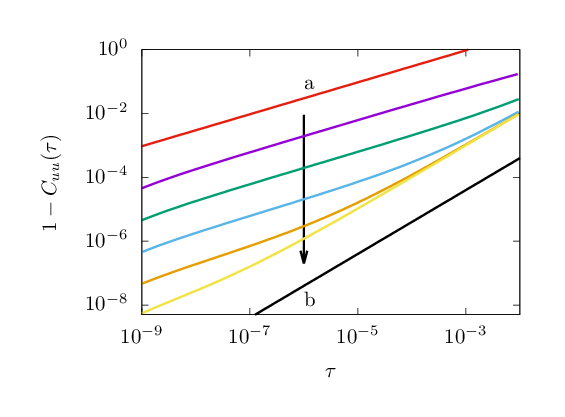}
\caption{Initial scaling of $1-C_{uu}^{(n)}(\tau)$ vs $\Delta \tau$
 of the solutions of the inertial
model eq. (\ref{eq9}) at  $\kappa_s=0$, $\Gamma=10^{10}$
for decreasing values of the prefactor $b$.
The arrow indicates decreasing values of $b=10^{0},\,10^{-1},\,10^{-2},\,
10^{-3},\, 10^{0}$. Line (a) represents the inertial scaling
 $1-C_{uu}^{(n)}(\tau)\sim \tau^{1/2}$, line (b) the Stokes scaling
$1-C_{uu}^{(n)}(\tau)\sim \tau$.}
\label{sFig2}
\end{figure}

Figure \ref{sFig2} depicts the short-time scaling 
of  $1-C_{uu}(\tau)$ (in nondimensional form) for decreasing
values of $b$. As $b$ decreases, the inertial scaling
$1-C_{uu}(\tau) \sim \tau^{1/2}$  occurs at smaller values of $\tau$,
followed  by a crossover to the Stokesian scaling,
$1-C_{uu}(\tau) \sim \tau$ , i.e. to the scaling controlled by
the Stokes friction in the absence of inertial effects.
For  example, at $b=10^{-2}$, the linear scaling
of   $1-C_{uu}(\tau)$ is evident starting from $\tau=10^{-3}$,
i.e.  for time scales a thousand times smaller than
the momentum relaxation time. This
explains why in the  experimental analysis of Duplat et al. \cite{duplat}
in gases, the authors found $m_{xx}(\tau) \sim \tau^3$ in the
range of their temporal resolution.

\section{H. Experimental data processing}
\label{smsec5}

The temporal resolution, $\tau_E$, is dictated by inherent noise sources in the experimental apparatus. When checking agreement between theory and experiment, it is necessary to account for experimental effects. We develop the techniques used to account for limited temporal resolution, the 8th-order finite differencing used to acquire the velocity time traces from position time traces, and the laser noise, which is relevant at high frequencies.

In the analyzed experimental position time traces of barium titanate glass in acetone, laser noise in the system limits the time resolution to 750 ns. We can account for this limited resolution directly in the theoretical position autocorrelation function. We write the resolution limited position variable $\bar{x}(t)$ in terms of $x(t)$ as

\begin{align}
    \bar{x}(t) = \frac{1}{\tau_E}\int_0^{\tau_E} x(t+s)ds
\end{align}

which leads to a resolution limited $\bar{C}_{xx}(\tau)$ of

\begin{multline}
    \bar{C}_{xx}(\tau) = \langle \bar{x}(t)\bar{x}(t+\tau) \rangle = \\
    \frac{1}{\tau_E^2} \langle \int_0^{\tau_E}x(t+s)ds\int_0^{\tau_E}x(t+\tau + \rho)d\rho\rangle .
\end{multline}

which we rewrite as

\begin{align}
    \bar{C}_{xx}(\tau) =\frac{1}{\tau_E^2} \int_0^{\tau_E}\int_0^{\tau_E}\langle x(t)x(t+\tau + \rho -s)\rangle dsd\rho,
\end{align}

using stationarity. With the resolution limited $\bar{C}_{xx}$ we can find the resolution limited velocity autocorrelation function $\bar{C}_{vv}$ using the exact finite differencing equation that is applied on the experimental position time trace to acquire a stable velocity time trace. The eighth order finite difference is written 

\begin{align}
    \bar{v}_{FD}(t) = \frac{1}{\tau_E}\sum_{n=-4}^{4} c_n \bar{x}(t+n\tau_E),
 \end{align}

where $c_n$ denotes the standard weights of $c_0 = 0$, $c_{\pm 1} = \pm 4/5$, $c_{\pm 2} = \mp 1/5$, $c_{\pm 3} = \pm 4/105$, and $c_{\pm 4} = \mp 1/280$.

We can find the expected binned and finite-differenced VACF as

\begin{align}
     \bar{C}_{vvFD}(\tau) = \frac{1}{\tau_E^2}\sum_{n=-4}^{4}\sum_{m=-4}^{4} c_n c_m \bar{C}_{xx}(|\tau+(m-n)\tau_E|).
     \label{binned_vacf_}
\end{align}

When comparing theoretical curves to velocity time series data we need to use $\bar{S}_v(\Delta \tau)$, which is found by replacing $C_{vv}(\tau)$ with $\bar{C}_{vv}(\tau)$ in eq. (\ref{eq_sda}). We then define a theoretical scaling factor $F(\Delta \tau)$ that deconvolves the artifacts of the signal processing by a simple ratio of $S_v(\Delta \tau)$ and $\bar{S}_v(\Delta \tau)$

\begin{equation}
    F(\Delta \tau) = \frac{S_v(\Delta \tau)}{\bar{S}_v(\Delta\tau)}.
    \label{map}
\end{equation}

Before we compare the processed theoretical curve to the experimental data, we also must account for laser noise effects. The experimentally measured $S_{v,E}(\Delta \tau)$ can be written as a combination of the Brownian signal and the laser noise

\begin{equation}
    \bar{S}_{v,E}(\Delta\tau) = E[|\Delta v_{BM}+\Delta v_{N}|]/\Delta\tau = \sqrt{\frac{2}{\pi}}\sqrt{\sigma_{BM}^2+\sigma_{N}^2}.
\end{equation}

Where the scale-dependent slope found from the laser noise alone is 

\begin{equation}
    \bar{S}_{v,N}(\Delta\tau) = E[|\Delta v_{N}|]/\Delta\tau = \sqrt{\frac{2}{\pi}}\sqrt{\sigma_{N}^2}.
\end{equation}

Therefore, we write the final Brownian scale-dependent slope in terms of the experimentally measured $\bar{S}_{v,E}(\Delta\tau)$ and the bright noise $\bar{S}_{v,N}(\Delta\tau)$ as 

\begin{equation}
    \bar{S}_{v,BM}(\Delta \tau) = \sqrt{\bar{S}_{v,E}(\Delta\tau)^2-\bar{S}_{v,N}(\Delta\tau)^2} 
    \label{brownian_signal}
\end{equation}

\begin{figure}
\includegraphics[width=8cm]{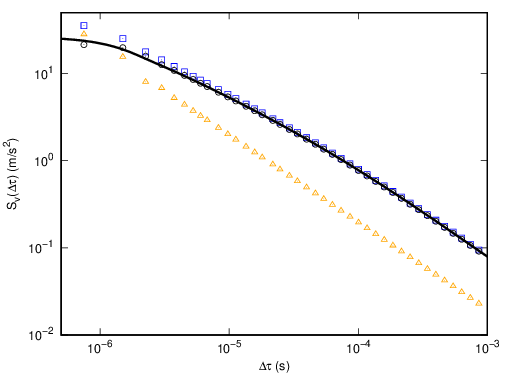}
\caption{Comparison of $\bar{S}_{v}(\Delta \tau)$ found for bright noise, experimental data, and resulting Brownian signal. The solid black curve represents the analytical theory for the finite-resolution, finite differenced $\bar{S}_{v}(\Delta \tau)$. The blue squares show $\bar{S}_{v,E}(\Delta\tau)$ found from experimental data, the yellow triangles show $\bar{S}_{v,N}(\Delta\tau)$ found from bright noise data, and the black circles show the resulting $\bar{S}_{v,BM}(\Delta \tau)$ found from equation (\ref{brownian_signal}).}
\label{sda_noise_incl}
\end{figure}

\begin{figure}
\includegraphics[width=8cm]{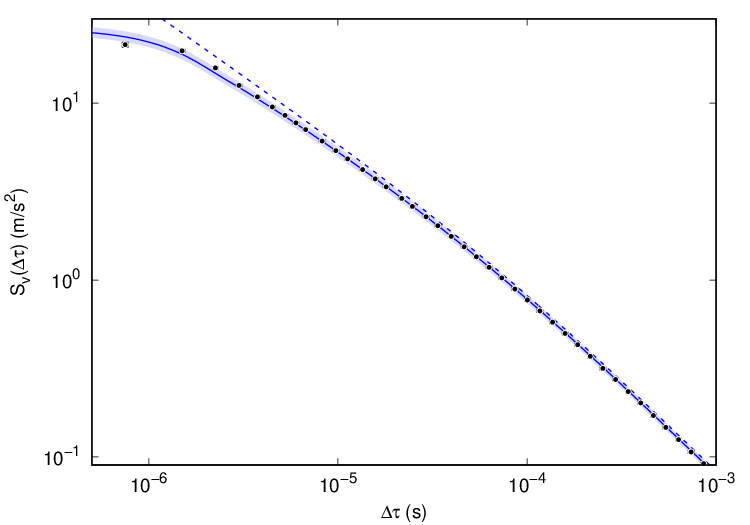}
\caption{Scale-dependent absolute slope $S_{v}(\Delta \tau)$ demonstrating time-resolution and finite differencing effects. The dashed blue curve $S_v(\Delta\tau)$ is the pure theory, which is explicitly convolved with the $750$ ns temporal binning and 8th-order finite-difference stencil to match the experimental signal processing pipeline, producing the solid blue theoretical curve $\bar{S}(\Delta\tau)$ which is compared to the data. The light blue envelope defines the $\pm 1\sigma$ parametric theoretical uncertainty determined by Monte Carlo propagation of trap and particle parameters. Statistical error bars on the data, computed via the blocking method and propagated through the noise subtraction are smaller than the marker size.}
\label{sda_adj}
\end{figure}

$\bar{S}_{v,BM}(\Delta \tau)$, $\bar{S}_{v,E}(\Delta\tau)$, and $\bar{S}_{v,N}(\Delta\tau)$ data are plotted together in figure \ref{sda_noise_incl} for comparison, along with the theoretical curve for $\bar{S}_{v,BM}(\Delta \tau)$. We can now compare the processed theory $\bar{S}_{v,BM}(\Delta \tau)$ to the data. We plot the processed and unprocessed theory (for reference), alongside the experimental data in figure \ref{sda_adj}. 

Finally we scale the time resolution limited data with $F(\Delta\tau)$ to create the final figure \ref{FigSDA} presented in the main text. In both figures, the statistical error bars are smaller than the plotted data points and are found via the blocking method \cite{flyvbjerg}. This method rigorously accounts for temporal correlations in the finite time series, with the variance propagated through the instrumental noise subtraction. Additionally, we propagate parametric uncertainty in particle radius, density, and trap stiffness \cite{btr} to the theoretical model via Monte Carlo sampling, where $\pm 1\sigma$ uncertainty intervals are represented as a light blue envelope in figure \ref{sda_adj}.

\section{I. Brownian universality class and fluctuation-dissipation
relations}
\label{smsec6}

In order to analyze the physical origin of the $5/2$-scaling
due to fluid inertia consider the modified Brownian dynamic
\begin{equation}
\begin{split}
\frac{d x(\tau)}{d \tau} &= u(\tau) \\
\frac{d u(\tau)}{d \tau} &= - u(\tau) - \kappa_s x(\tau) - G u(\tau) \\
&\quad + \sum_{k=1}^{N_i} \gamma_k \mu_k z_k(\tau) + \sqrt{2} \xi(\tau) \\
\frac{d z_k(\tau)}{d \tau} &= - \mu_k z_k(\tau) + u(\tau)
\end{split}
\label{equ1}
\end{equation}

that corresponds to the Generalized Langevin Equation
\begin{equation}
\frac{d u(\tau)}{d \tau} = - \int_0^\tau k(\tau-\theta)
\left ( \frac{d u(\theta)}{d \theta} + u(0) \delta (\theta) \right )
\,  d \theta- \kappa_s x(\theta) + R(\theta)
\label{equ2}
\end{equation}
with  $k(\tau)=  \sum_{k=1}^{N_i} \gamma_k \, e^{-\mu_k \, \tau}= b \, g(\tau,\gamma,N_i)$,
and  the stochastic forcing $R(\tau)$ given by
\begin{equation}
R(\tau) = \sqrt{2}  \, \xi(\tau) +\sum_{k=1}^{N_i} \mu_k \, \gamma_k z_{k,0}
\, e^{-\mu_k \, \tau}
\label{equ3}
\end{equation}
where $z_{k,0}$ are Gaussian random variables.
Apart from this exponentially decaying term with time,  $R(\tau)$ is
proportional to the distributional derivative of a Wiener process,
and eq. (\ref{equ1}) does not satisfy the Kubo fluctuation-dissipation
relation.
Eq. (\ref{equ1}) corresponds to the fluid-inertial
dynamics of a particle forced by a stochastic Wiener-noise
contribution.
\begin{figure}
\includegraphics[width=8cm]{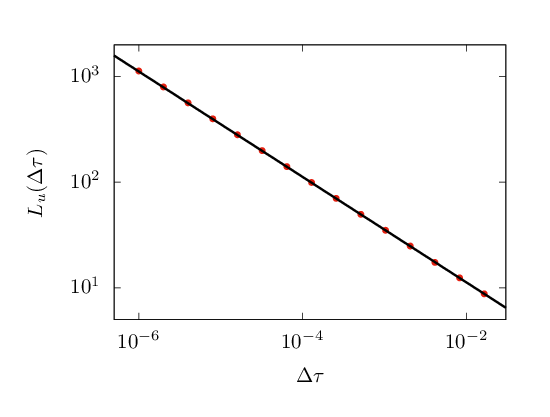}
\caption{Length-resolution analysis $L_u(\Delta \tau)$ vs $\Delta \tau$
(symbols 
$\bullet$)
of a velocity realization of eq. (\ref{equ1}) at $\kappa_s=0.1$, $b=0.1$,
$\Gamma=10^8$. The solid line represents the scaling $L_u(\Delta \tau)
\sim \Delta \tau^{-1/2}$.}
\label{sFig3}
\end{figure}
Figure \ref{sFig3} depicts the scaling of $L_u(\Delta \tau)$ vs $\Delta \tau$,
corresponding in this case to a pure Wiener-scaling characterized by
the exponent $1/2$.

This  result enables us to conclude that the
fractal dimension $d_v=7/4$, or equivalently the H\"older exponent $H_v=1/4$
in Brownian motion in liquids is the combined
effect of the fluid-inertial hydromechanic contribution coupled,
at thermal equilibrium, to the model for the stochastic force $R(\tau)$
consistent with the fluctuation-dissipation relations. This
essentially means that the short-time scaling  of
Brownian motion depends on the specific scaling
properties of the stochastic thermal-hydrodynamic force
$R(\tau)$  associated with fluid inertial effects,
and  deriving from fluctuation-dissipation relations  at thermal
equilibrium.

\end{document}